\documentclass[prl,twocolumn,tbtags,showpacs,superscriptaddress,floatfix]{revtex4}

\usepackage{amsmath}
\usepackage{graphicx}

\begin{document}

\title{Conductance characteristics of current-carrying \textit{d}-wave weak
links}
\author{S.N. Shevchenko}
\affiliation{B. Verkin Institute for Low Temperature Physics and Engineering, 47 Lenin
Ave., 61103, Kharkov, Ukraine}
\date{\today}

\begin{abstract}
The local quasiparticle density of states in the current-carrying \textit{d}%
-wave superconducting structures was studied theoretically. The density of
states can be accessed through the conductance of the scanning tunnelling
microscope. Two particular situations were considered: the current state of
the homogeneous film and the weak link between two current-carrying \textit{d%
}-wave superconductors.
\end{abstract}

\maketitle

\section{Introduction}

Unconventional superconductors exhibit different features interesting both
from the fundamental point of view and for possible applications \cite%
{review(d-wave)}. In particular, double degenerated state can be realized in
\textit{d}-wave Josephson junctions \cite{review(KOZ)}. If the
misorientation angle between the banks of the junction $\chi $ is taken $\pi
/4$, the energy minima of the system appear at the order parameter phase
difference $\phi =\pm \pi /2$. These degenerate states correspond to the
counter flowing currents along the junction boundary. Such characteristics
make \textit{d}-wave Josephson junctions interesting for applications, such
as qubits \cite{dd-qubit}. Our proposition was to make these qubits
controllable with the externally injected along the boundary transport
current \cite{KOSh04}. It was shown that the transport current and the
spontaneous one do not add up -- more complicated interference of the
condensate wave functions takes place. This is related to the phenomena,
known as the paramagnetic Meissner effect \cite{review(d-wave)}.

It was demonstrated both experimentally \cite{Braunisch} and theoretically
\cite{Higashitani, FRS97} that at the boundary of some high-$T_{c}$
superconductors placed in external magnetic field the current flows in the
direction opposite to the diamagnetic Meissner supercurrent which screens
the external magnetic field. This countercurrent is carried by the
surface-induced quasiparticle states. These nonthermal quasiparticles appear
because of the sign change of the order parameter along the reflected
quasiparticle trajectory. Such a depairing mechanism is absent in the
homogeneous situation. Note that in a homogeneous conventional
superconductor at zero temperature the quasiparticles appear only when the
Landau criterion is violated, at $v_{s}>\Delta _{0}/p_{F}$. Here $v_{s}$\ is
the superfluid velocity which parameterizes the current-carrying state, $%
\Delta _{0}$ stands for the bulk order parameter, and $p_{F}$\ is the Fermi
momentum. The appearance of the countercurrent can be understood as the
response of the weak link with negative self-inductance to the externally
injected transport supercurrent. The state of the junction in the absence of
the transport supercurrent at zero temperature is unstable at $\phi =\pi $\
from the point of view that small deviations $\delta \phi =\pm 0$\ change
the Josephson current\ from $0$\ to its maximal value \cite{KO}. The
response of the Josephson junction to small transport supercurrent at $\phi
=\pi $\ produces the countercurrent \cite{Sh04}. It is similar to the
equilibrium state with the persistent current in 1D normal metal ring with
strong spin-orbit interaction: there is degeneracy at zero temperature\ and $%
\phi =\pi $, and the response of the ring is different at $\delta \phi \neq
0 $\ or $B\neq 0$, where $B$\ is the effective magnetic field which enters
in the Hamiltonian through the Zeeman term (which breaks time-reversal
symmetry) \cite{1DNring}. The degeneracy is lifted by small effective
magnetic field so that the persistent current rapidly changes from $0$\ to
its maximum value. In the case of the weak link between two superconductors
in the absence of the transport supercurrent there is degeneracy between $%
+p_{y}$ and $-p_{y}$\ zero-energy states; both the time-reversal symmetry
breaking by the surface (interface) order-parameter and the Doppler shift
(due to the transport supercurrent or magnetic field) lift the degeneracy
and result in the surface (interface) current \cite{Braunisch}.

In recent years mesoscopic superconducting structures continue to attract
attention because of the possible application as qubits, quantum detectors
etc. (e.g. \cite{dd-qubit}, \cite{Jtransistor}). In particular, such
structures can be controlled by the transport supercurrent and the magnetic
flux (through the phase difference on Josephson contact). This was in the
focus of many recent publications, e.g. \cite{KOSh04, Rashedi04, Zhang04,
Ngai05, Sh06, Lukic07}. Here we continue to study the mesoscopic
current-carrying \textit{d}-wave structures. Particularly, we study the
impact of the transport supercurrent on the density of states in both
homogeneous film and in the film which contains a weak link.

\section{Model and basic equations}

We consider a perfect contact between two clean singlet superconductors. The
external order parameter phase difference $\phi $ is assumed to drop at the
contact plane at $x=0$. The homogeneous supercurrent flows in the banks of
the contact along the $y$-axis, parallel to the boundary. The sample is
assumed to be smaller than the London penetration depth so that the
externally injected transport supercurrent can indeed be treated as
homogeneous far from the weak link. The size of the weak link is assumed to
be smaller than the coherence length. Such a system can be quantitatively
described by the Eilenberger equation \cite{KO}. Taking transport
supercurrent into account leads to the Doppler shift of the energy variable
by $\mathbf{p}_{F}\mathbf{v}_{s}$. The standard procedure of matching the
solutions of the bulk Eilenberger equations at the boundary gives the
Matsubara Green's function $\widehat{G}_{\omega }(0)$ at the contact at $x=0$
\cite{KOSh04}. Then for the component $G_{\omega }^{11}\equiv g(\omega ,%
\mathbf{r})$ of $\widehat{G}_{\omega }$, which defines both the current
density and the density of states (see below), we obtain in the left ($L$)
and right ($R$) banks of the junction:%
\begin{equation}
g_{L},_{R}(\mathbf{r})=g_{L},_{R}(\mathbf{\infty })+\left[
g(0)-g_{L},_{R}(\infty )\right] \exp \left( -\frac{2\left\vert \mathbf{r}%
\right\vert \Omega _{L,R}}{\left\vert v_{x}\right\vert }\right) ,
\label{g_L_R}
\end{equation}%
\begin{equation}
g_{L},_{R}(\infty )=\frac{\widetilde{\omega }}{\Omega _{L,R}},  \label{g_inf}
\end{equation}%
\begin{equation}
g(0)=\frac{\widetilde{\omega }(\Omega _{L}+\Omega _{R})-i\text{sgn}%
(v_{x})\Delta _{L}\Delta _{R}\sin \phi }{\Omega _{L}\Omega _{R}+\widetilde{%
\omega }^{2}+\Delta _{L}\Delta _{R}\cos \phi }.  \label{g(0)}
\end{equation}%
Here $\omega =\pi T(2n+1)$ are Matsubara frequencies, $\Delta _{L,R}$ stands
for the order parameter in the left (right) bank, and
\begin{equation}
\widetilde{\omega }=\omega +i\mathbf{p}_{F}\mathbf{v}_{s},\text{ \ \ }\Omega
_{L,R}=\sqrt{\widetilde{\omega }^{2}+\Delta _{L,R}^{2}}.
\end{equation}%
The direction-dependent Doppler shift $\mathbf{p}_{F}\mathbf{v}_{s}$ results
in the modification of current-phase dependencies and in the appearance of
the countercurrent along the boundary.

The function $g(\omega ,\mathbf{r})$ defines the current density, as
following:
\begin{equation}
\mathbf{j}=4\pi eN_{0}v_{F}T\sum\limits_{\omega _{n}>0}\left\langle \widehat{%
\mathbf{v}}\text{Im}g\right\rangle _{\widehat{\mathbf{v}}}.  \label{j}
\end{equation}%
Here $N_{0}$ is the density of states at the Fermi level, $\left\langle
...\right\rangle _{\widehat{\mathbf{v}}}$ denotes averaging over the
directions of Fermi velocity $\mathbf{v}_{F}$, $\widehat{\mathbf{v}}=\mathbf{%
v}_{F}/v_{F}$ is the unit vector in the direction of $\mathbf{v}_{F}$.

Analytic continuation of $g(\omega )$, \textit{i.e.}
\begin{equation}
g(\varepsilon )=g(\omega \rightarrow -i\varepsilon +\gamma ),  \label{g(e)}
\end{equation}%
gives the retarded Green's function, which defines the density of states:
\begin{equation}
N(\varepsilon ,\mathbf{r})=\text{Re}g(\varepsilon ,\mathbf{r}).  \label{N}
\end{equation}%
Here $\gamma $ is the relaxation rate in the excitation spectrum of the
superconductor.

The local density of states can be probed with the method of the tunnelling
spectroscopy by measuring the tunnelling conductance $G=dI/dV$ of the
contact between our superconducting structure and the normal metal scanning
tunnelling microscope's (STM) tip. At low temperature the dependence of the
conductance on the bias voltage $V$ is given by the following relation \cite%
{Duke}:%
\begin{equation}
G(eV)=G_{N}\left\langle D(\mathbf{p}_{F})N(eV,\mathbf{p}_{F})\right\rangle ,
\label{G(eV)}
\end{equation}%
where $G_{N}$ is the conductance in the normal state; $D(\mathbf{p}_{F})$ is
the angle-dependent superconductor-insulator-normal metal barrier
transmission probability. The barrier can be modelled \textit{e.g. }as in
Ref. \cite{FRS97} with the uniform probability within the acceptance cone $%
\left\vert \vartheta \right\vert <\vartheta _{c}$, where $\vartheta $ is the
polar angle and the small value of $\vartheta _{c}$ describes the thick
tunnelling barrier:%
\begin{equation}
D(\vartheta )=\frac{1}{2\vartheta _{c}}\theta (\vartheta _{c}^{2}-\vartheta
^{2}),  \label{D}
\end{equation}%
where $\theta (...)$ is the theta function.

\section{Conductance characteristics of the homogeneous current-carrying film%
}

\begin{figure}[h]
\includegraphics[width=8cm]{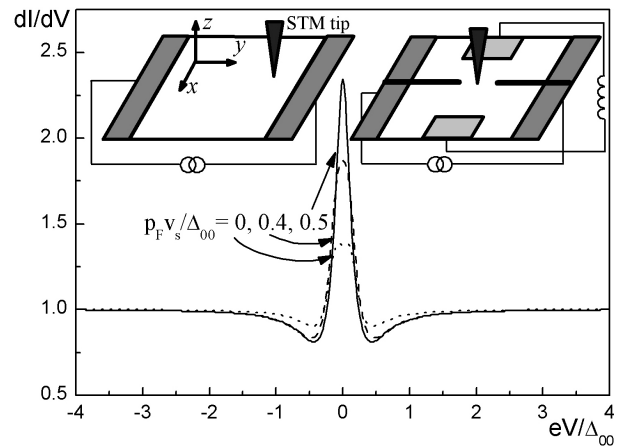}
\caption{Normalized (divided by $G_{N}$) conductance $dI/dV$ for the
homogeneous current-carrying state in the \textit{d}-wave film for different
values of the transport current. The curves are plotted with $\protect\gamma %
/\Delta _{00}=0.15$ and $\protect\vartheta _{c}=0.1\protect\pi $ [$\Delta
_{00}=\Delta _{0}(v_{s}=0)$]. Left and right insets show the schemes for
probing the density of states in the current-carrying \textit{d}-wave film
and in the weak link (see text for details). }
\label{dIdV}
\end{figure}

Before studying the current-carrying weak link we consider the homogeneous
situation. We will consider the \textit{d}-wave film as shown in the left
inset in Fig. \ref{dIdV}. The motivation behind this study is twofold:
first, to demonstrate the application of the theory presented above, and
second, to describe recent experimental results \cite{Ngai05}.

The system considered consists of the \textit{d}-wave film, in which the
current is injected along the $y$-axis, and the STM normal metal tip
(another STM contact is not shown in the scheme for simplicity; for details
see \cite{Ngai05}). Following the experimental work \cite{Ngai05}, we
consider the $c$-axis along the $x$-axis and the misorientation angle
between $a$-axis and the direction of current ($y$-axis) to be $\pi /4$.
Such problem can be described with the equations presented in the previous
section as following \cite{FRS97, Sh06}.

Consider the specular reflection at the border, when the boundary between
the current-carrying $d$-wave superconductor and the insulator can be
modelled as the contact between two superconductors with the order
parameters given by $\Delta _{L}=\Delta (\vartheta )=\Delta _{0}\cos
2(\vartheta -\chi )$ and $\Delta _{R}=\Delta (-\vartheta )\equiv \overline{%
\Delta }$ and with $\phi =0$. Then from Eq. (\ref{g(0)}) we have the
following:%
\begin{equation}
g(\omega )=\frac{\widetilde{\omega }\left( \Omega +\overline{\Omega }\right)
}{\Omega \overline{\Omega }+\widetilde{\omega }^{2}+\Delta \overline{\Delta }%
},  \label{g_for_d-film}
\end{equation}%
where $\Omega =\sqrt{\widetilde{\omega }^{2}+\Delta ^{2}}$ and $\overline{%
\Omega }=\sqrt{\widetilde{\omega }^{2}+\overline{\Delta }^{2}}$. This
expression is valid for any relative angle $\chi $ between the $a$-axis and
the normal to the boundary; in particular,%
\begin{eqnarray}
g(\omega ) &=&\frac{\widetilde{\omega }}{\Omega },\text{ \ }\chi =0\text{ \
\ }(\Delta (\vartheta )=\Delta _{0}\cos 2\vartheta ), \\
g(\omega ) &=&\frac{\Omega }{\widetilde{\omega }},\text{ \ }\chi =\frac{\pi
}{4}\text{ \ }(\Delta (\vartheta )=\Delta _{0}\sin 2\vartheta ).
\label{g0_chi=pi4}
\end{eqnarray}%
The accurate dependence of the gap function $\Delta _{0}=\Delta
_{0}(v_{s},\gamma )$ can be obtained from Ref. \cite{KOSh04} with
introducing $\gamma $ as following: $\mathbf{p}_{F}\mathbf{v}%
_{s}\longrightarrow \mathbf{p}_{F}\mathbf{v}_{s}-i\gamma $ (which is
analogous to Eq. (\ref{g(e)})). The energy values in this paper are made
dimensionless with the zero-temperature gap at zero current: $\Delta
_{00}=\Delta _{0}(v_{s}=0)$.

And now with Eqs. (\ref{g0_chi=pi4}) and (\ref{g(e)}-\ref{G(eV)}) we plot
the STM conductance for the current-carrying $d$-wave film in Fig. (\ref%
{dIdV}). We obtain the suppression of the zero-bias conductance peak by the
transport supercurrent, as was studied in much detail in Ref. \cite{Ngai05}.
Our results are in agreement with their Fig. 1. Also the authors of Ref.
\cite{Ngai05} developed the model based on phase fluctuations in the BTK
formalism to explain the suppression of the zero-bias conductance peak.
However, their theoretical result, Fig. 2, describes the experimental one
only qualitatively, leaving several distinctions. They are the following:
(i) position of the minima ($eV/\Delta _{00}\sim 0.5$ and $1$ for the
experiment and the theory respectively); (ii) height of the zero-bias peak
at zero transport current ($\sim 2.5$ and $4$ respectively); (iii) height of
the peak at maximal transport current ($\sim 1.3$ and $2.5$ respectively);
(iv) presence/absence of the minima for all curves. Our calculations, Fig. (%
\ref{dIdV}), demonstrate agreement with the experiment in all these
features. The agreement we obtained with two fitting parameters, $\gamma $
and $\vartheta _{c}$.
\begin{figure}[h]
\includegraphics[width=8cm]{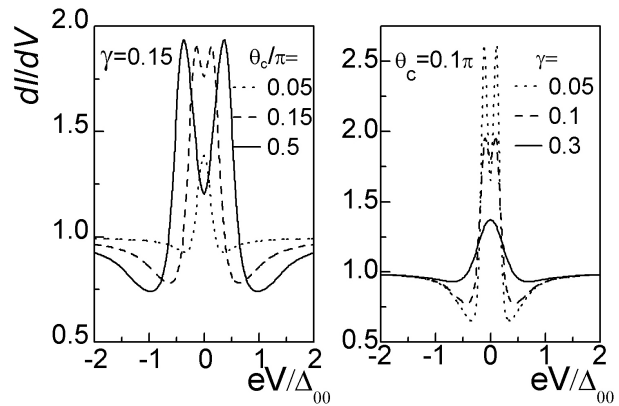}
\caption{Normalized conductance $dI/dV$ for the homogeneous current-carrying
state in the \textit{d}-wave film for different values of $\protect\gamma $
and $\protect\vartheta _{c}$ at $p_{F}v_{s}/\Delta _{00}=0.5$.}
\label{splitting}
\end{figure}

To further demonstrate the impact of the two fitting parameters of our
model, $\gamma $ and $\vartheta _{c}$, in Fig. (\ref{splitting}) we plot the
normalized conductance fixing one of them and changing another. The figure
clearly demonstrates how they change the shape of the curves: the position
of the minima, splitting of the zero-bias peak etc. Note that the splitting
is suppressed at small $\vartheta _{c}$ and high $\gamma $. This absence of
the splitting was observed in the experiment \cite{Ngai05} and studied in
several articles, e.g. \cite{Aubin}.

\section{Conductance characteristics of the current-carrying weak link}

Consider now the weak link between two $d$-wave current-carrying banks. For
studying the effect of both the transport current and the phase difference
on the density of states in the contact, we propose the scheme, presented in
the right inset in Fig. \ref{dIdV}. The supercurrent is injected along $y$%
-axis in the superconducting film, as it was discussed in the previous
section. Besides, the weak link is created by the impenetrable for electrons
partition at $x=0$. The small break in this partition ($a<\xi _{0}$) plays
the role of the weak link in the form of the pinhole model \cite{KO,
FogYipKurk, Sh06}. The STM tip in the scheme is positioned above the weak
link to probe the density of states in it. Two more contacts along the $x$%
-axis provide the order parameter phase difference $\phi $ along the weak
link. This can be done, for example, by connecting the contacts with the
inductance, as shown in the scheme, and applying magnetic flux $\Phi _{e}$
to this inductance. Then one obtains the phase control of the contact with
the relation: $\phi =\Phi _{e}/\Phi _{0}$.

\begin{figure}[h]
\includegraphics[width=8cm]{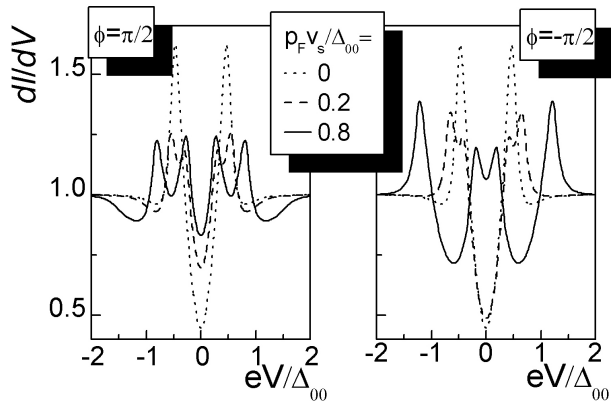}
\caption{Normalized conductance $dI/dV$ at the contact between two
current-carrying $d$-wave superconductors for different values of the
transport supercurrent ($v_{s}$) and for two values of the phase difference $%
\protect\phi =\pm \protect\pi /2$.}
\label{conductance}
\end{figure}

The two half-plains (for $x<0$ and $x>0$) play the role of the two banks of
the contact, which we also call left and right superconductors. In our
scheme the banks carry the transport current along the boundary, and the
Josephson current along the contact is created due to the phase difference.
The banks we consider to be $d$-wave superconductors with $c$-axis along the
$z$-axis and with the misorientation angles $\chi _{L}=0$ and $\chi _{R}=\pi
/4$. Now we can apply the equations presented in Sec. II to describe the
conductance characteristics of the contact between current-carrying $d$-wave
superconductors. This is done in Fig. \ref{conductance}, where the
normalized conductance is plotted for two values of the phase difference,
for $\phi =\pm \pi /2$ and with $\gamma /\Delta _{00}=0.1$. The two values
of the phase difference, $\phi =\pm \pi /2$, are particularly interesting
for the application since they correspond to the double-degenerate states
\cite{dd-qubit, KOSh04}. So, the density of states is the same in the
absence of the transport supercurrent in both panels in Fig. \ref%
{conductance} with mid-gap states (at $eV<\Delta _{00}$) which create the
spontaneous current along the boundary. The transport supercurrent ($%
v_{s}\neq 0$) removes the degeneracy by significantly changing the mid-gap
states (Fig. \ref{conductance}), which explains different dependencies of
the current in the contact on the applied transport current (i.e. on $v_{s}$%
), studied in \cite{KOSh04, Sh04}.

\section{Conclusion}

We have studied the density of states in the current-carrying $d$-wave
structures. Namely, we have considered, first, the homogeneous situation
and, second, the superconducting film with the weak link. The former case
was related to recent experimental work, while the latter is the proposition
for the new one. The local density of states was assumed to be probed with
the scanning tunnelling microscope. The density of states at the weak link
and the current (\textit{i.e.} its components through the contact and along
the contact plane) are controlled by the values of $\phi $\ and $v_{s}$. \
The system is interesting because of possible applications: in the Josephson
transistor with controlling parameters $\phi $\ and $v_{s}$\ governed by
external magnetic flux and the transport supercurrent \cite{Jtransistor},
and in solid-state qubits, based on a contact of $d$-wave superconductors
\cite{dd-qubit}.

\begin{acknowledgments}
The author is grateful to Yu.A. Kolesnichenko and A.N. Omelyanchouk for
helpful discussions. This work was supported in part by the Fundamental
Researches State Fund (grant number F28.2/019).
\end{acknowledgments}

\end{document}